\documentclass[aps,physrev,twocolumn,groupedaddress]{revtex4-2}

\usepackage{graphicx}
\usepackage{siunitx}
\usepackage{amsmath}
\usepackage{csquotes}
\usepackage{hyperref}
\hypersetup{colorlinks=true,linkcolor=blue,citecolor=blue,urlcolor=blue}

\newcommand*{\figref}[2][]{%
  \hyperref[{fig:#2}]{%
    \ref*{fig:#2}%
    \ifx\\#1\\%
    \else
      #1%
    \fi
  }%
}

\newcommand{\CC}{C\nolinebreak\hspace{-.05em}\raisebox{.4ex}{\tiny\bf +}\nolinebreak\hspace{-.10em}\raisebox{.4ex}{\tiny\bf +}}
\def\CC{{C\nolinebreak[4]\hspace{-.05em}\raisebox{.4ex}{\tiny\bf ++}}}

\begin{document}

\title{Dependence of charge separation efficiency on the exciton--charge transfer offset and Gaussian disorder in organic solar cells}

\author{Maik Schwuchow}
\email[Contact author: ]{maik.schwuchow@physik.tu-chemnitz.de}
\author{Carsten Deibel}
\author{Angela Thränhardt}
\affiliation{Institute of Physics, Chemnitz University of Technology, 09107 Chemnitz, Germany}

\date{\today}

\begin{abstract}

State-of-the-art organic solar cells increasingly rely on low-offset semiconductor blends, challenging the traditional requirement of a large energetic driving force for efficient charge separation. In these systems, the energetic offset $\Delta E_{\mathrm{LE-CT}}$ between local exciton (LE) and charge-transfer (CT) states approaches the thermal energy, making exciton--CT hybridization and thermal repopulation of the exciton level critical to device performance. In this work, we directly compare a macroscopic two-state rate model with three-dimensional kinetic Monte--Carlo (kMC) simulations to investigate microscopic charge separation dynamics and the role of Gaussian energetic disorder. We demonstrate that in the absence of disorder, the analytical rate model accurately reproduces kMC predictions for the whole range of $\Delta E_{\mathrm{LE-CT}}$. Specifically, the macroscopic model successfully explains horizontal shifts in the internal quantum efficiency curves that arise depending on how the energetic offset is physically realized in the constituent molecules. We show that these variations can be captured entirely through the ratio of degeneracies of the LE and CT states, respectively. Introducing Gaussian energetic disorder into the kMC simulation reveals a distinct crossover behavior depending on $\Delta E_{\mathrm{LE-CT}}$. While disorder is mostly detrimental at large offsets, it  can significantly boost efficiency at intermediate and low offsets. Thermalization of charge carriers within the disorder-broadened density of states creates an effective driving force allowing charge separation even at zero or negative energetic offsets.

\end{abstract}

\maketitle

\section{Introduction}

Organic solar cell (OSC) research, using both experimental and theoretical approaches, has been thriving for multiple decades at this point \cite{antolinOrganicSolarCells2025}. State-of-the-art cells have reached power conversion efficiencies of over \SI{20}{\percent} \cite{chenOrganicSolarCells2025,antolinOrganicSolarCells2025}, increasingly closing the gap to their inorganic counterparts. Nowadays, OSCs typically employ low-offset material blends, where the energetic differences of the frontier molecular orbitals between acceptor and donor material are minimized \cite{bertrandieEnergyLevelConundrum2022,classenRoleExcitonLifetime2020,hanLocalExcitationChargeTransfer2019,liMappingEnergyLevel2022,liuFastChargeSeparation2016,perdigon-toroExcitonsDominateEmission2021}. Although the specific values reported in literature can vary, e.\,g., due to differences in measurement techniques \cite{bertrandieEnergyLevelConundrum2022,liMappingEnergyLevel2022,neusserSpectroelectrochemicallyDeterminedEnergy2022}, the general notion of low offsets stands in stark contrast to the common belief of a large energetic driving force being a necessary requirement for efficient charge separation \cite{balzerDelocalisationEnablesEfficient2024,liuFastChargeSeparation2016,rileyRoleExcitonDiffusion2022,scharberEfficiencyBulkheterojunctionOrganic2013}. Low-offset blends seem to circumvent this requirement, significantly reducing energy losses and thereby increasing open-circuit voltages \cite{eisnerHybridizationLocalExciton2019,fritschInterplayCTSinglet2022a,balzerDelocalisationEnablesEfficient2024,liuFastChargeSeparation2016}. So it is by no means surprising, that the dissociation of optically generated intramolecular excitons into separated charge carriers is still at the heart of OSC research \cite{zhuSmallExcitonBinding2021,lowrieOrganicPhotovoltaicsCurrent2023,liangEfficientHoleTransfer2021,liuFastChargeSeparation2016}. Especially the properties and intricacies of charge-transfer (CT) states that are located at donor--acceptor interfaces and constitute the intermediate state between exciton and separated charges attract much attention \cite{kahleStaticDynamicDisorder2022,khanInfluenceDisorderState2021,khanQuantifyingEffectEnergetic2022,hussainRoleInterfaceExtraction2021,gaspariniAdjustingEnergyInterfacial2021,eisnerHybridizationLocalExciton2019,liuManagingChallengesOrganic2022,perdigon-toroBarrierlessFreeCharge2020,gohlerRoleDynamicStatic2022,yanInfluenceStaticDisorder2021}. One example in this regard is thermalization, i.\,e., the loss of non-equilibrium excess energy \cite{athanasopoulosDisorderVsDelocalization2019,saladinaChargePhotogenerationNonFullerene2021,upretiSlowRelaxationPhotogenerated2021,wojcikElucidatingRoleDisorder2022}. Significant evidence points in the direction that non-thermalized charge carriers, especially \emph{hot} CT states at donor/acceptor interfaces, facilitate charge separation in OSCs \cite{saladinaChargePhotogenerationNonFullerene2021,jonesAreHotCharge2014,upretiSlowRelaxationPhotogenerated2021,vijiHotCarrierOrganic2024,wojcikElucidatingRoleDisorder2022}. Then again, also separation that primarily proceeds through thermalized CT states has been discussed in the literature \cite{athanasopoulosDisorderVsDelocalization2019,BasslerHotorcold2015}. In recent years, with the rise of low-offset blends, exciton--CT hybridization has shifted into focus more and more \cite{eisnerHybridizationLocalExciton2019,hanLocalExcitationChargeTransfer2019, balzerDelocalisationEnablesEfficient2024,gillettRoleChargeRecombination2021}. As the energetic offset $\Delta E_{\mathrm{LE-CT}}$ between the local exciton (LE) and the interfacial CT state approaches the thermal energy, a strict differentiation between the states might not be reasonable/possible anymore \cite{eisnerHybridizationLocalExciton2019,balzerDelocalisationEnablesEfficient2024,xuTuningHybridizationLocal2020}. Furthermore, an efficient charge back-transfer from the CT to the excitonic state and therefore a dynamic equilibrium between the two populations has to be considered in such a scenario \cite{classenRoleExcitonLifetime2020,gillettRoleChargeRecombination2021,fritschInterplayCTSinglet2022a,perdigon-toroExcitonsDominateEmission2021}. This thermal exciton repopulation helps avoid CT recombination losses by increasing the relative exciton population. Since the lifetimes of CT states are usually much higher than the ones of excitons, this will automatically yield higher excitonic losses, though \cite{hoodEntropyDisorderEnable2016}. Accordingly, the beneficial effect of high exciton lifetimes has been emphasized by multiple authors \cite{classenRoleExcitonLifetime2020,rileyRoleExcitonDiffusion2022}. This scenario only gets more complicated when including disorder effects \cite{gohlerRoleDynamicStatic2022,perdigon-toroUnderstandingRoleOrder2022,yanInfluenceStaticDisorder2021}. The local energetic landscape \cite{saladinaPowerLawDensityOfStates2023}, often approximated by static Gaussian disorder \cite{basslerChargeTransportDisordered1993,felekidisAutomatedOpensourceSoftware2018,pasveerTemperatureChargeCarrier2005,wojcikChargeTransportDisordered2020}, significantly influences the charge separation dynamics \cite{athanasopoulosDisorderVsDelocalization2019,athanasopoulosWhatisthebindingenergy2019,hoodEntropyDisorderEnable2016,hussainRoleInterfaceExtraction2021}. Therefore, in this work, we want to investigate the influence of energetic disorder on charge separation efficiency in low-offset systems.
\begin{figure*}
\includegraphics[width=1\textwidth]{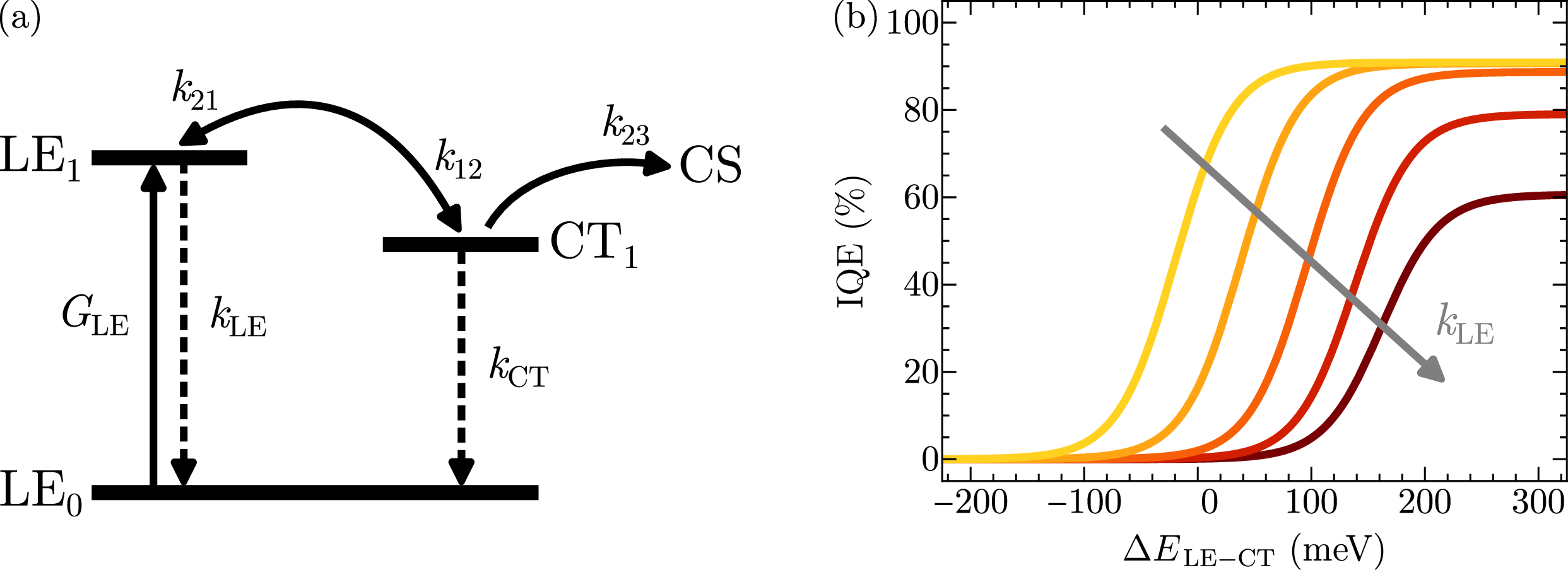}
\caption{(a) Jablonski diagram of the two-state rate model illustrating the local exciton (LE) and CT states as well as the transition and decay rate constants. (b) Exemplary plots of Eq.\,\ref{IQErateSolution} for an increase of the exciton decay rate $k_{\mathrm{LE}}$ of about 4 orders of magnitude. All other parameters are constant.}
\label{fig:RateModelFig}
\end{figure*}

Modeling OSCs is not an easy task. Relevant processes and phenomena can span multiple orders of magnitude of both space and time \cite{kongUnderstandingNonradiativeCharge2025}. The inherently statistical nature of, e.\,g., hopping transport in general and disorder in particular, often renders analytical approaches difficult and less versatile. Depending on the specific scenario one is interested in, different models and simulation techniques should be considered \cite{casalegnoNumericalSimulationPhotocurrent2013,casalegnoCoarsegrainedKineticModelling2012}. In this regard, kinetic Monte-Carlo (kMC) simulations are the gold standard for modeling OSCs. They are highly customizable and can cover a plethora of microscopic effects. They usually consist of a lattice of localization sites with particles hopping between them. By simulating individual particle trajectories using pseudo-random numbers, this comes at the cost of computational resources, though. Over the years, the kMC method has been used to study the influence of, e.\,g., morphology \cite{casalegnoMethodologicalAssessmentKinetic2010a, heiberImpactTortuosityChargeCarrier2017}, energetic disorder \cite{athanasopoulosDisorderVsDelocalization2019,basslerChargeTransportDisordered1993,kaiserKineticMonteCarlo2019}, localization/delocalization \cite{athanasopoulosDisorderVsDelocalization2019,athanasopoulosWhatisthebindingenergy2019,balzerDelocalisationEnablesEfficient2024,saladinaChargePhotogenerationNonFullerene2021} and particle interactions/recombination \cite{heiberEncounterLimitedChargeCarrierRecombination2015,heiberChargeCarrierConcentration2016,gillettRoleChargeRecombination2021,SaxenaKineticMonteCarloStudy2020}. While plenty of sophisticated simulations of OSCs have been reported in the literature, the majority of them focuses on the traditional high-offset point of view \cite{kaiserGeneralizedKineticMonte2018a,poppImpactPhosphorescentSensitizers2019}. This disregards the influence low $\Delta E_{\mathrm{LE-CT}}$ can have. Analytical investigations, on the other hand, usually have to rely on the modeling of a limited number of effects using effective-medium or mean-field approximations \cite{fishchukUnifiedDescriptionHopping2013,casalegnoNumericalSimulationPhotocurrent2013}. While keeping these limitations in mind, e.\,g., master equation and drift--diffusion models have been used extensively in the last decades \cite{casalegnoNumericalSimulationPhotocurrent2013,cuppenKineticMonteCarlo2013,fallahpourModelingSimulationEnergetically2014,gagliardiSimulationChargeCarrier2018,pasveerTemperatureChargeCarrier2005}. Furthermore, even minimal analytical models can prove useful in explaining experimental results. Notably, when ignoring all morphological and dimensional aspects, a set of two first-order rate equations describing the populations of excitons and CT states is able to capture the steady-state LE--CT equilibrium observed at low offsets \cite{classenRoleExcitonLifetime2020}.

In order to bridge the gap between the macroscopic abstractions of analytical equations and the microscopic, stochastic nature of kMC simulations, we will use both models in this work. We will connect the predictions of a steady-state rate model directly to three-dimensional kMC data with the main parameter being the energetic offset $\Delta E_{\mathrm{LE-CT}}$. By initially ignoring disorder effects, we validate the employed kMC implementation, demonstrating that both models can produce nearly identical results. Moreover, we illustrate the parallels between different realizations of $\Delta E_{\mathrm{LE-CT}}$ in the simulation to the degeneracy ratio of the states in the rate model. From there, including Gaussian disorder in the simulation reveals the limits of the rate equation as both detrimental and beneficial effects appear in the kMC data depending on $\Delta E_{\mathrm{LE-CT}}$. We start by explaining the model details in section \ref{sec2}. Section \ref{sec3} covers an in-depth discussion of the obtained results.

\section{Model details\label{sec2}}

\subsection{Two state rate model}

Arguably one of the simplest analytical approaches to describe OSC behavior is solving a set of coupled rate equations in the steady-state limit \cite{langentepekongDarkSideOfOrganic2026}. For the present case, we limit the model to two species: local excitons (LE) and CT states as done by Classen et al. \cite{classenRoleExcitonLifetime2020}. They are to be understood as singlet states, ignoring triplets in this work. By assigning constant rates to all major processes involving these states, as illustrated in Fig.\,\figref[(a)]{RateModelFig}, the matrix form of the rate model is easily written down,
\begin{equation}
		\begin{pmatrix}
			\frac{\mathrm{d}}{\mathrm{d}t} n_\mathrm{LE} \\
			\frac{\mathrm{d}}{\mathrm{d}t} n_\mathrm{CT} \\
		\end{pmatrix}
		=
		\begin{pmatrix}
			G_{\mathrm{LE}}-k_{\mathrm{LE}} & k_{\mathrm{21}} \\
			k_{\mathrm{12}} & -k_{\mathrm{CT}}-k_{\mathrm{23}}
		\end{pmatrix}
		\begin{pmatrix}
			n_\mathrm{LE} \\
			n_\mathrm{CT}
		\end{pmatrix}.
	\label{rateEQ}
\end{equation}
Here, $n_{\mathrm{LE}}$ and $n_{\mathrm{CT}}$ are the LE and CT densities, respectively. $G_{\mathrm{LE}}$ is the LE generation rate corresponding to photon absorption, $k_{\mathrm{12}}$ is the transfer rate from LE to CT, $k_{\mathrm{21}}$ is the transfer rate from CT to LE, $k_{\mathrm{LE}}$ and $k_{\mathrm{CT}}$ are the LE and CT decay rates. There is no distinction between radiative and non-radiative decay, i.\,e., $k_{\mathrm{LE}}$ and $k_{\mathrm{CT}}$ are to be understood as total decay rates. As long as reabsorption of emitted photons is not modeled/relevant, this does not constitute an approximation as radiative and non-radiative rates would always appear together. Finally, the rate $k_{\mathrm{23}}$ describes the splitting of CT states into free charges which are promptly extracted and therefore do not constitute an additional third species of particles. Consequently, the IQE is given by
\begin{equation}
	\mathrm{IQE}=\frac{n_{\mathrm{CT}}\, k_{\mathrm{23}}}{G_{\mathrm{LE}}}.
	\label{IQErateEQ1}		
\end{equation}
Given the constant LE generation rate $G_{\mathrm{LE}}$, the LE and CT populations will approach a dynamical (quasi-) equilibrium. The corresponding steady-state populations can be calculated analytically from Eq.\,\ref{rateEQ} by setting the derivatives on the left-hand side to zero and solving the system of linear equations. Substituting the steady-state solution for $n_{\mathrm{CT}}$ into Eq.\,\ref{IQErateEQ1} yields the steady-steady IQE:
\begin{equation}
	\mathrm{IQE}=\left(1+\frac{k_{\mathrm{CT}}}{k_{\mathrm{23}}}+\frac{k_{\mathrm{LE}}}{k_{\mathrm{12}}}\left(1+\frac{k_{\mathrm{21}}}{k_{\mathrm{23}}}+\frac{k_{\mathrm{CT}}}{k_{\mathrm{23}}}\right)\right)^{-1}.
	\label{IQErateSolution}		
\end{equation}
Approximating the LE--CT quasi-equilibrium in a classical way using the Boltzmann distribution, the probability $p_i$ of either state being occupied varies exponentially with the normalized energy $E_i/k_{\mathrm{B}}T$ ($k_{\mathrm{B}}$ is the Boltzmann constant, $T$ is the absolute temperature) and depends linearly on its degeneracy $g_i$:
\begin{equation}
	p_i \propto g_i \exp\left(-\frac{E_i}{k_\mathrm{B} T}\right), \quad i=\mathrm{LE},\, \mathrm{CT}.
\end{equation}
Furthermore, in the quasi-equilibrium case, the ratio of probabilities $p_{\mathrm{CT}}/p_{\mathrm{LE}}$ is equal to the ratio of transfer rates $k_{\mathrm{21}}/k_{\mathrm{12}}$, yielding
\begin{equation}
		\frac{k_{\mathrm{21}}}{k_{\mathrm{12}}}=\frac{p_{\mathrm{CT}}}{p_{\mathrm{LE}}}=\frac{g_{\mathrm{CT}}}{g_{\mathrm{LE}}} \exp \left(\frac{\Delta E_{\mathrm{LE-CT}}}{k_{\mathrm{B}}T}\right),
	\label{probratio}
\end{equation}
with $\Delta E_{\mathrm{LE-CT}}=E_{\mathrm{LE}}-E_{\mathrm{CT}}$. Fig.\,\figref[(b)]{RateModelFig} shows five plots of Eq.\,\ref{IQErateSolution} with $k_{\mathrm{21}}$ calculated using Eq.\,\ref{probratio}. Depending on the rates used the characteristic sigmoid curves will differ in height and onset of the plateau region at positive offsets $\Delta E_{\mathrm{LE-CT}}$. 

\subsection{Kinetic Monte-Carlo simulation}

\begin{figure}
\includegraphics[width=0.483\textwidth]{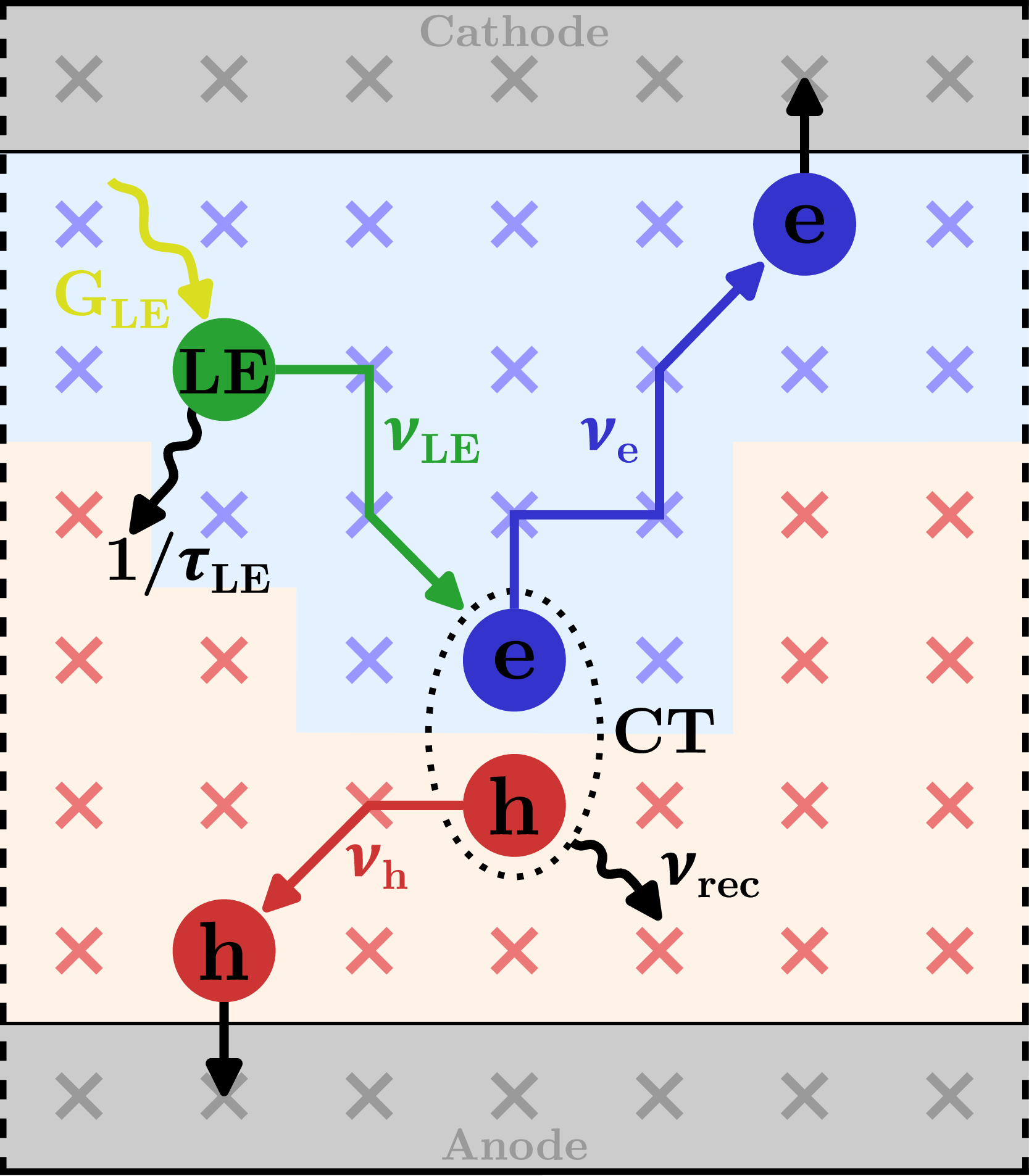}
\caption{Simplified 2D schematic illustrating the main aspects of charge separation in the kinetic Monte-Carlo simulation used in this work. Local exciton (LE) generation occurs with a constant rate on a simple cubic lattice of donor (red) and acceptor (blue) sites. Excitons move diffusively without disorder effecting the exciton energy levels. In order to separate the charges, the excitons have to dissociate into a charge transfer (CT) state at a donor--acceptor interface. If the charge carriers can overcome their mutual Coulomb attraction and move toward their respective electrodes (grey), an extraction event can occur. Excitons and CT states may recombine to the ground state, removing them from the simulation. In the two directions parallel to the electrodes, periodic boundary conditions apply.}
\label{fig:SimSchematic}
\end{figure}
In order to study low-offset OSCs in larger detail, we employed a kinetic Monte-Carlo simulation (kMC) written in \CC. In the following, we outline the main aspects of our implementation. The main processes and steps are schematically shown in Fig.\,\figref{SimSchematic}. A three-dimensional simple cubic lattice of zero-dimensional sites is generated inside a simulation box with edge-length $L$. Throughout this work, we set $L=\SI{100}{\nano\m}$ and the lattice constant $l=\SI{1}{\nano\m}$. All particle species considered in the simulation are localized on individual sites. One pair of opposing sides is defined to be adjacent to metallic contacts, i.\,e., cathode and anode. In the two directions parallel to these contacts, periodic boundary conditions are applied. Then, each site has to be set to either represent a donor or acceptor molecule. In order to separate charge carriers efficiently, donor and acceptor sites cannot be randomly distributed in the simulation box \cite{cuiMorphologyOptimizationPhotoactive2021}. Instead, a bulk-heterojunction morphology, i.\,e., a bi-continuous network of two phases, resembling the active layer of a real cell, is created by the Ising-based algorithm of Heiber \cite{heiberEfficientGenerationModel2014, heiberIsing_OPVV40Experimental2018}. The donor--acceptor ratio was set to \num{0.5} and the other parameters were chosen to create average domain sizes of about \SI{10}{\nano\m}. To represent electron and hole transport layers, the two layers of sites adjacent to the cathode and anode are defined as donor- and acceptor-type, respectively. Sites are characterized by the energies $E_{\mathrm{HOMO}}$ and $E_{\mathrm{LUMO}}$, indicating the highest occupied and lowest unoccupied molecular orbitals, respectively. The baseline offsets between the two materials, i.\,e., $E_{\mathrm{HOMO,don}}-E_{\mathrm{HOMO,acc}}$ and $E_{\mathrm{LUMO,don}}-E_{\mathrm{LUMO,acc}}$, are set to \SI{800}{\milli\eV} each. As described below, these offsets will be varied in order to model a low-offset blend. In order to describe energetic disorder, using a single (pseudo-) random number for each site, a static energy shift is drawn from a Gaussian distribution with mean zero and standard deviation $\sigma_{\mathrm{dis}}$. We assume that local electrostatic interactions influence all energy levels the same way \cite{balzerDelocalisationEnablesEfficient2024}. Therefore, both LUMO and HOMO energy of a site are shifted by the same value, representing perfectly correlated disorder \cite{balzerDelocalisationEnablesEfficient2024}.

Exciton generation is modeled using a constant generation rate of $G_{\mathrm{LE}}=\SI{0.1}{\per\cubic\nano\m\per\s}$ \footnote{High disorder can lead to charge trapping which results in vanishing hop rates. In this scenario, exciton generation becomes more likely, increasing the number of particles present in the simulation. At $\sigma_{\mathrm{dis}}=\SI{125}{\milli\eV}$, we used $G_{\mathrm{LE}}=\SI{0.01}{\per\cubic\nano\m\per\s}$ for this reason. This has no influence on the results.}. This value is one order of magnitude lower than typical for constant illumination by unfocused sunlight in order to ensure the absence of non-geminate interactions \cite{casalegnoMethodologicalAssessmentKinetic2010a,poppImpactPhosphorescentSensitizers2019}. In other words, in the present work, we only study the separation dynamics of individual excitons, i.\,e., individual electron--hole pairs. We note that homogeneous generation of excitons disregards interference effects as well as differing absorption coefficients of donor and acceptor molecules \cite{arminSpectralDependenceInternal2014}. Excitons are considered to be electrically neutral. Furthermore, only singlet excitons with an average lifetime $\tau_{\mathrm{LE}}$ are considered. Exciton decay is modeled using a constant rate of $1/\tau_{\mathrm{LE}}=k_{\mathrm{LE}}$, which corresponds to the total rate of all radiative and non-radiative pathways. Exciton decay results in the removal of the exciton from the simulation box and does not include potential reabsorption of emitted photons. Exciton transport on the other hand is described as hopping, i.\,e., (thermally assisted) tunneling between sites. The rates $\nu_{\mathrm{LE}}$ for exciton jumps are calculated using the formula for F\"orster resonance energy transfer (FRET) corresponding to singlet exciton transport \cite{forsterZwischenmolekulareEnergiewanderungUnd1948,kaiserGeneralizedKineticMonte2018a}:
\begin{equation}
    \nu_{\mathrm{LE}}(r)=\nu_{0,\mathrm{LE}}\ r^{-6}\exp\left(-\frac{\Delta E + \vert\Delta E\vert}{2k_{\mathrm{B}}T}\right).
    \label{forster}
\end{equation}
Here, $\Delta E$ is the energy difference between final and initial state, $r$ is the intersite distance, $T=\SI{293}{\kelvin}$ is absolute temperature and $\nu_{0,\mathrm{LE}}=\SI{e11}{\per\s}$. $\nu_{0,\mathrm{LE}}$ is constant throughout this work and incorporates the F\"orster radius and the donor excitations total decay rate \cite{kaiserGeneralizedKineticMonte2018a}. We note that this simplifies the dipole--dipole interaction, which FRET is based on, by assuming perfectly aligned dipole moments and disregarding molecular orientations. Furthermore, we disregard possible exciton--exciton interaction. $\nu_{0,\mathrm{LE}}$ is chosen arbitrarily in order to allow a sufficiently high percentage of excitons to reach the donor--acceptor interface at the given domain size.

Excitons are modeled as composite particles, comprising one electron and one hole each. Therefore, also the rates for electron and hole jumps have to be considered at every timestep. Hopping rates of charged particles $\nu_{\mathrm{e/h}}$ are calculated using a Miller--Abrahams \cite{millerImpurityConductionLow1960} type equation:
\begin{equation}
    \nu_{\mathrm{e/h}}(r,\,\Delta E)=\nu_{0,\mathrm{e/h}}\,\exp\left(-\frac{2r}{\alpha}\right)\exp\left(-\frac{\Delta E + \vert\Delta E\vert}{2k_{\mathrm{B}}T}\right),
    \label{MArates}
\end{equation}
with the constant prefactor ${\nu_{0,\mathrm{e/h}}=\SI{5e12}{\per\s}}$ and the localization parameter ${\alpha=\SI{0.5}{\nano\m}}$. The interaction energy of an electron and a hole on the same site, i.\,e., when they form an exciton, is called the binding energy ${E_{\mathrm{B,LE}}=\SI{700}{\milli\eV}}$. Therefore, the local exciton level is defined as ${E_{\mathrm{LE}}=E_{\mathrm{LUMO}}-E_{\mathrm{HOMO}}-E_{\mathrm{B,LE}}}$. Although \SI{700}{\milli\eV} is significantly higher than reported experimental values in state-of-the-art materials \cite{neusserSpectroelectrochemicallyDeterminedEnergy2022,zhuSmallExcitonBinding2021}, a high $E_{\mathrm{B,LE}}$ in the simulation is necessary in order to restrict exciton dissociation to donor--acceptor interfaces. When the exciton binding energy approaches the CT binding energy defined below, local excitons will no longer be stable and dissociation in the pure phases becomes significant.

Interactions between charges on different sites are calculated using Coulomb's law with a homogeneous dielectric constant of $\varepsilon_{\mathrm{r}}=\num{3.5}$ \cite{casalegnoMethodologicalAssessmentKinetic2010a}. We note that charges near a conductor (or a different dielectric medium) will induce charges on the surface of the conductor. This can be modeled using the method of image charges, which are known to influence charge dynamics \cite{casalegnoMethodologicalAssessmentKinetic2010a}. Including such images in the simulation and investigating their influence on charge separation in low-offset systems exceeds the scope of this work and will be done in a later publication. In the present case, we restrict the implementation of image charges to the two directions parallel to the contacts using the minimum-image convention \cite{casalegnoMethodologicalAssessmentKinetic2010a}. This corresponds to including all interactions with every other particle exactly once, either via the original or its first periodic replica.

If an exciton reaches an interfacial site during its lifetime, the energy difference $\Delta E_{\mathrm{LE-CT}}$ between exciton and charge-transfer (CT) state is critical. A CT state is occupied if either the electron of an exciton in the donor phase jumps to an acceptor site or if the hole of an exciton in the acceptor phase jumps to a donor site. Either way, a CT state is comprised of an electron in the acceptor phase and a hole in the donor phase. The rates for these jumps are calculated with Eq.\,\ref{MArates}. For this exciton dissociation to occur and be stable, the offset
\begin{align}
    \Delta E_{\mathrm{LE-CT}}&=E_{\mathrm{LE}}-E_{\mathrm{CT}}\nonumber\\
    &=E_{\mathrm{LE}}-(E_{\mathrm{LUMO,acc}}-E_{\mathrm{HOMO,don}}-E_{\mathrm{B,CT}})
    \label{offset}
\end{align}
has to be favorable, i.\,e., sufficiently positive. $E_{\mathrm{B,CT}}$ is the CT binding energy. The simplest interpretation of $E_{\mathrm{B,CT}}$ is as the Coulomb interaction energy of the (initial) electron--hole pair. We note that the simulation is not explicitly limited to dissociation events involving only nearest-neighbor sites. In the on-lattice model we use, the most probable initial separation distance is one lattice constant, though. For this reason, we set $E_{\mathrm{B,CT}}$ in all calculations of $\Delta E_{\mathrm{LE-CT}}$ to be equal to the Coulomb interaction energy of an electron and a hole at a distance of \SI{1}{\nano\m}. For $\varepsilon_{\mathrm{r}}=\num{3.5}$, this yields $E_{\mathrm{B,CT}}\approx\SI{411}{\milli\eV}$. In accordance with the discussion of the exciton binding energy before, this value of $E_{\mathrm{B,CT}}$ is higher than reported values in the literature \cite{KrohAnImpedenceStudy2024}. The discrepancy can be readily explained by larger initial CT separations depending on the specific molecules used and, e.\,g., delocalization and thermalization of the charge carriers \cite{saladinaChargePhotogenerationNonFullerene2021,athanasopoulosDisorderVsDelocalization2019}. Then again, since the main parameter in this study is $\Delta E_{\mathrm{LE-CT}}$, shifting both the excitonic and CT binding energies will not change the qualitative results presented below.

CT states may recombine to the ground state. The rate for this process is calculated using
\begin{equation}
    \nu_{\mathrm{rec}}=\nu_{\mathrm{0,rec}}\,\exp\left(-\frac{2r}{\alpha}\right),
    \label{CTrecomb}
\end{equation}
with the constant prefactor $\nu_{\mathrm{0,rec}}=\SI{5e7}{\per\s}$. Since CT recombination to the ground state is the major loss mechanism at high offsets, we chose this value of $\nu_{\mathrm{0,rec}}$ in order to guarantee a realistically high IQE in this offset regime. On the other hand, CT states may also recombine to a local exciton state in either phase. This exciton repopulation is treated as the inverse process of exciton separation into a CT state with rates modeled by Eq.\,\ref{MArates}. While recombination to the ground state is inherently energetically favorable, exciton repopulation depends explicitly on $\Delta E_{\mathrm{LE-CT}}$.

\begin{figure}
\includegraphics[width=0.483\textwidth]{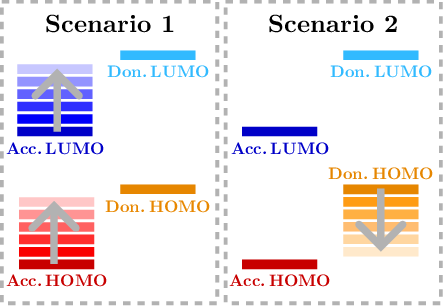}
\caption{Diagram illustrating the two scenarios of varying $\Delta E_{\mathrm{LE-CT}}$ in the simulation. Scenario 1 increases HOMO and LUMO energies of the acceptor in equal steps, keeping the donor energies constant. Scenario 2 decreases the HOMO energy of the donor, keeping the acceptor energies constant. In both cases, the energy of the CT state at a donor--acceptor interface increases (see Eq.\,\ref{offset}). Scenario 2 also increases the donor exciton energy.}
\label{fig:ScenarioSchematic}
\end{figure}

The variation of $\Delta E_{\mathrm{LE-CT}}$ in the simulation is done by varying the HOMO and LUMO energies, respectively. We differentiate two scenarios, illustrated in Fig.\,\figref{ScenarioSchematic}. In scenario 1, both LUMO and HOMO of the acceptor material are increased by the same amount. The donor energy levels stay fixed. This keeps $E_{\mathrm{LE,acc}}=E_{\mathrm{LE,don}}$ constant, increasing only $E_{\mathrm{CT}}$. Scenario 2 keeps the acceptor energy levels constant, while the HOMO of the donor is decreased. The local exciton level in the donor phase $E_{\mathrm{LE,don}}$ hereby increases, as well as $E_{\mathrm{CT}}$. Because $E_{\mathrm{LE,don}}$ changes in scenario 2, we use $E_{\mathrm{LE,acc}}$ for the calculation of $\Delta E_{\mathrm{LE-CT}}$ in this case. This is justified because excitons will accumulate in the lower energy phase, which therefore dominates the subsequent behavior. An experimental scenario similar to scenario 2 has already been studied in the literature \cite{classenRoleExcitonLifetime2020}. We note that, while scenario 1 treats electrons and holes symmetrically, exciton repopulation in scenario 2 is mainly driven by holes.

Charge carriers that approach a contact can be extracted. The contacts are modeled as an additional layer of sites, respectively. A jump to a contact site extracts the charge carrier immediately, removing it from the simulation. The rate for this jump is calculated using Eq.\,\ref{MArates} and is assumed to be energetically favorable for both electrons and holes, i.\,e., $\Delta E\leq0$. Thermal injection, i.\,e., hopping of charge carriers from the Fermi level of the electrode into the active layer, is not included in the simulation. This simplification is justified because we are only interested in the short-circuit condition and will not simulate voltage sweeps. A linear potential gradient is assumed between the contacts with a built-in voltage of \SI{1}{\volt}. This means that charged particles on all sites on a plane parallel to the contacts have the same potential energy.

In each timestep, all possible rates for processes involving sites in a cube of edge-length \SI{6}{\nano\m} centered on each individual particle are calculated. The \num{26} highest rates for each particle are considered. This definition allows the description of both isoenergetic transport, where nearest-neighbor jumps are dominant, and hopping in a disordered landscape, where local energetic minima might require longer distance jumps. Choosing a process and calculating the timestep is done using the BKL algorithm \cite{bortzNewAlgorithmMonte1975}. This means that the rates of all possible events are summed up and the resulting $\nu_{\mathrm{tot}}$ is interpreted as one over the average waiting time between events. As is typical for a Poisson process, this time is distributed exponentially. Using a randomly chosen real number $x \in (0,1]$, the timestep $\Delta t$ is calculated as
\begin{equation}
    \Delta t = -\frac{1}{\nu_{\mathrm{tot}}}\,\ln(x).
\end{equation}
Finally, the partial sums of all possible rates (normalized with $\nu_{\mathrm{tot}}$) are calculated, and a process is chosen using a second random number. In other words, the probability of a specific process occurring in this step is equal to its rate divided by $\nu_{\mathrm{tot}}$.
\begin{figure*}
\includegraphics[width=1\textwidth]{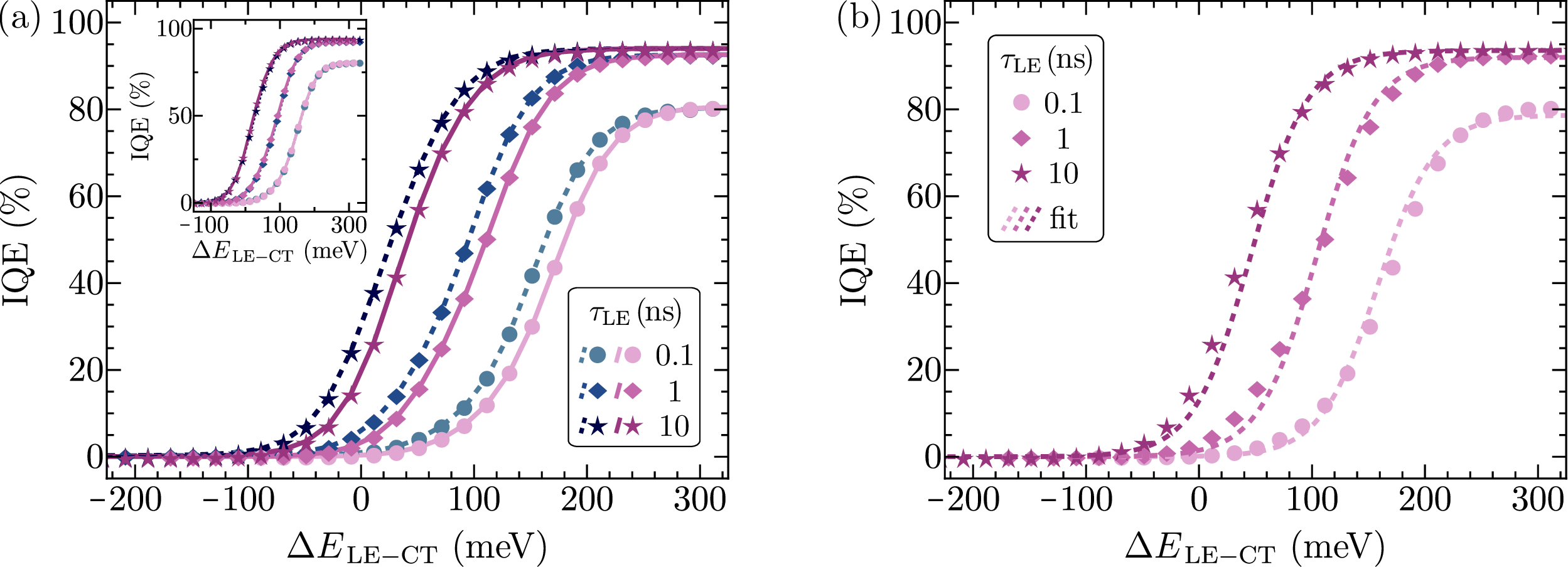}
\caption{(a) kMC data on the dependence of the internal quantum efficiency (IQE) on the energetic offset of local exciton (LE) and charge-transfer state (CT) for three exciton lifetimes $\tau_{\mathrm{LE}}$. Solid lines and red tones represent scenario 1 (varying acceptor HOMO and LUMO), dashed lines and blue tones represent scenario 2 (varying donor HOMO) of creating the offset in the simulation (see Fig.\,\figref{ScenarioSchematic}). The sigmoidal shape from rate model theory is clearly visible. As the offset decreases, exciton repopulation and decay increase, which limits the IQE. The inset shows, that the data for both scenarios coincides, when a horizontal shift is applied (see text). All lines (dashed and solid) are merely guides for the eye. (b) The scenario 1 data from (a) is plotted including a fit of the rate model. Only the $\tau_{\mathrm{LE}}=\SI{1}{\nano\s}$ data has been fitted explicitly. The upper/lower curves use the fitted model except $k_{\mathrm{LE}}$ decreased/increased by an order of magnitude, respectively. The accuracy of the fit is not perfect, which is expected due to fundamental differences between the dynamics that are describable by the two models.}
\label{fig:tauresults}
\end{figure*}
One simulation run for a specific parameter set consists of \num{e5} exciton generations. Our main figures of merit are the average number of electron and hole extractions, $n_{\mathrm{ext}}$ and $p_{\mathrm{ext}}$, at their respective contacts. This corresponds to the steady-state limit of the rate model discussed before. Results are averaged over \num{8} runs, each with an individual morphology. Because non-geminate effects are negligible in the low-density limit studied here, the internal quantum efficiency
\begin{equation}
    \mathrm{IQE}=\frac{1}{2}\frac{n_{\mathrm{ext}}+p_{\mathrm{ext}}}{\num{e5}}
\end{equation}
is equal to the exciton separation efficiency at the donor--acceptor interfaces. In order to gain more insight into the underlying physics of charge separation, we may also plot the average numbers of exciton and CT recombinations, respectively. These are a direct measure of the relative steady-state exciton and CT populations. In fact, disregarding electron extraction at the anode and hole extraction at the cathode, i.\,e., losses by contact recombination, the sum of the relative exciton population, CT population and IQE will equal \SI{100}{\percent} in our simulation.

\section{Results and discussion\label{sec3}}

\subsection{Isoenergetic ordered system}

Comparing solutions of the rate model and the kMC simulation is a necessary step in order to validate the implementation of the latter as well as outline the limits of the former. Monte-Carlo calculations require large amounts of CPU-time, whereas plotting Eq.\,\ref{IQErateSolution} is a matter of seconds. So as to not waste resources, the limits of the rate model should be well understood. Because the rate model inherently ignores disorder effects, we compare it with the zero-disorder kMC. The rate model steady-state IQE is pictured in Fig.\,\figref[(b)]{RateModelFig}, whereas a similar plot is shown in Fig.\,\figref[(a)]{tauresults} for the kMC for both scenarios (see Fig.\,\figref{ScenarioSchematic}) and three values of the exciton lifetime $\tau_{\mathrm{LE}}=1/k_{\mathrm{LE}}$. Varying $\Delta E_{\mathrm{LE-CT}}$ yields a sigmoid shape of the IQE in both models. The difference between the two scenarios of creating $\Delta E_{\mathrm{LE-CT}}$ in the kMC only comes down to a horizontal shift of the IQE curves. The inset of Fig.\,\figref[(a)]{tauresults} shows that the data coincides, if the scenario 1 data is shifted by about $\SI{-17.5}{\milli\eV}$ along the horizontal axis. Fig.\,\figref[(b)]{tauresults} shows the scenario 1 data from Fig.\,\figref[(a)]{tauresults} as well as a fit of the rate model to the data points. We note that only the $\tau_{\mathrm{LE}}=\SI{1}{\nano\s}$ data has been fitted explicitly. By holding all other fitted parameters constant, only increasing and decreasing $k_{\mathrm{LE}}$ by an order of magnitude, respectively, the other two dashed lines emerge.
\begin{figure*}
\includegraphics[width=1\textwidth]{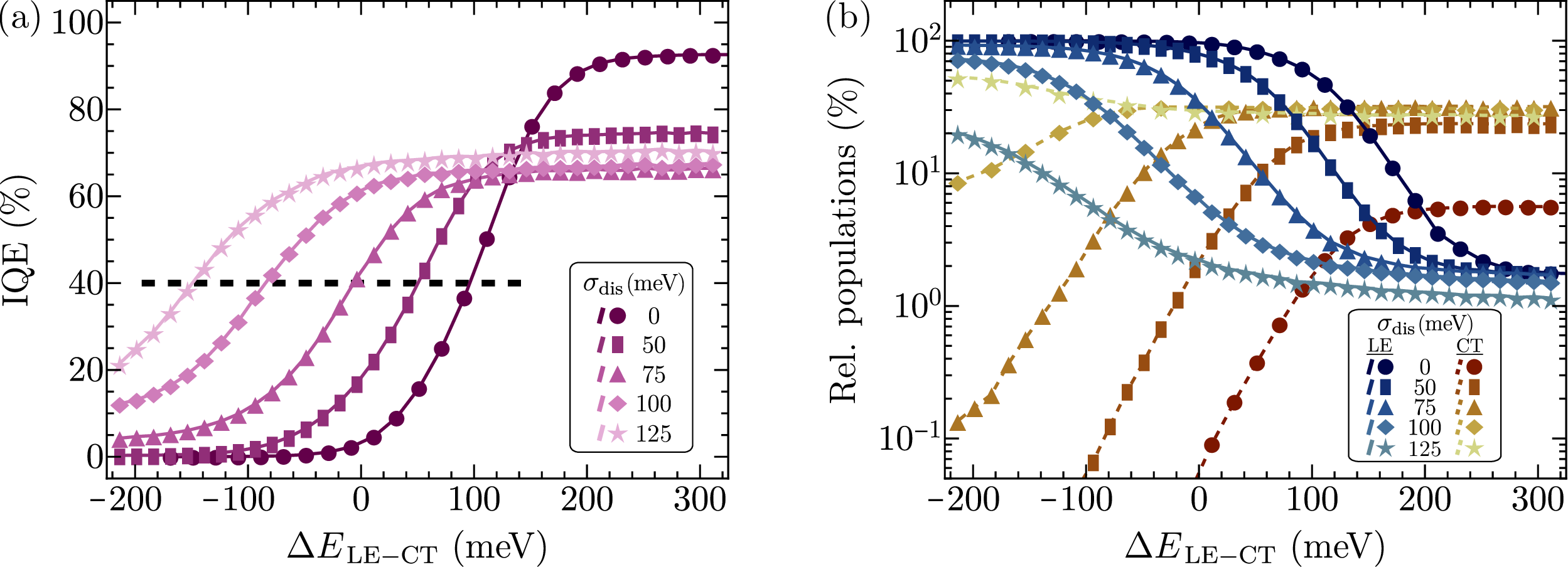}
\caption{(a) kMC steady-state data of the internal quantum efficiency (IQE) and (b) relative population dependence on the energetic offset between local exciton (LE) and charge transfer (CT) state $\Delta E_{\mathrm{LE-CT}}$ at various disorder strengths characterized by the standard deviation $\sigma_{\mathrm{dis}}$. $\tau_{\mathrm{LE}}$ has been set to \SI{1}{\nano\s}. In order to create the offset, scenario 1 of the energy level shift as described in the main text and Fig.\,\figref{ScenarioSchematic} was used in the simulation. Introducing disorder in the classic high-offset regime reproduces the mostly detrimental trend for charge separation known in the literature (see main text), that arises from trapping and reduced carrier mobilities. Low-offset systems can greatly benefit from energetic disorder. A key aspect is the repopulation of excitonic states from CT states. The observed IQE enhancement originates in the avoidance of exciton recombination due to energetic relaxation in the disordered density of states. The dashed line in (a) marks a cross-section of the IQE data at \SI{40}{\percent}, plotted in Fig.\,\figref[(b)]{disLinFit} as a function of $\sigma_{\mathrm{dis}}$. The percentage was chosen in order to try to guarantee a symmetrical cut through all data sets. All lines (solid and dashed) are merely guides to the eye.}
\label{fig:disresults}
\end{figure*}

Qualitatively, the main difference between the high and low $\Delta E_{\mathrm{LE-CT}}$ systems is the existence of a stable CT state. The ratio of relative populations of local exciton and CT state shift continuously to the exciton side as the offset decreases. This is illustrated for the kMC simulation in Fig.\,\figref[(b)]{disresults}. The direct consequence is a steep decline of the IQE. This does not mean that the charge carriers are harder to separate at low offsets, but rather that the CT states are not present long enough for a separation to occur. Both models reproduce these general trends of the population dynamics and IQE. Furthermore, the rate model is able to reproduce the disorder-free kMC results to a high degree. As is obvious from Eq.\,\ref{IQErateSolution}, the rate model solutions are not unique though. E.\,g., a constant factor multiplied with all rates cancels out completely. Moreover, increasing the parameter space to a three-state model, i.\,e., including free charges explicitly, does not yield higher quality fits in the present case (not shown) \footnote{Since the kMC simulation in this work is setup to focus on the separation of single electron--hole pairs only, higher order effects like non-geminate recombination are not significant. By changing this premise, including additional species in the rate model will become necessary eventually. E.\,g., a sophisticated 5-state rate model has been studied recently by Langentepe-Kong et al. \cite{langentepekongDarkSideOfOrganic2026}.}. On the contrary, even the two-state model is over-determined in that rates of the model can be set arbitrarily if the others are adjusted accordingly. Nevertheless, the resulting fits are not perfect, which is to be expected. The mathematics of the rate model and the kMC appear to be identical at first sight, but the underlying dynamics of charge motion are more intricate than a system of two linear equations can recreate. In fact, the first analytical description of geminate recombination, given by Onsager \cite{onsagerInitialRecombinationIons1938}, uses a drift-diffusion differential equation. Though he assumed the case of a vanishing reaction distance, this model has been extended to include more realistic boundary conditions \cite{tachiyaGeneralMethodCalculating1978,sanoPartiallyDiffusioncontrolledRecombination1979,sekiDispersivediffusioncontrolledDistancedependentRecombination2006}. It has also been shown that hopping models are equivalent to continuous diffusion models, when the hop lengths are short enough \cite{wojcikGeminateElectronholeRecombination2010}. Either way, the bottom line is that diffusive motion under the influence of an attractive field, e.\,g., a CT state separating into free charges, is not a simple first-order rate process \cite{wojcikGeminateElectronholeRecombination2017,wojcikAccuraciesEmpiricalTheories2009}. 

Keeping this limitation in mind, the rate model is still able to qualitatively and quantitatively explain the observed horizontal shift between the two scenarios of the kMC in Fig.\,\figref[(a)]{tauresults}. We already mentioned that scenario 2, i.\,e., the HOMO shift in the donor, not only increases $E_{\mathrm{CT}}$ but also $E_{\mathrm{LE,don}}$. This increase of $E_{\mathrm{LE,don}}$ leads to an accumulation of excitons in the acceptor phase, because jumps upward in energy are unfavorable (see Eq.\,\ref{forster}). The decay energy statistics of excitons supports this by exhibiting a dominant signal from the lower-$E_{\mathrm{LE}}$ phase (not shown). In other words, because the donor--acceptor ratio is \num{0.5}, half of the sites, i.\,e., half of the excitonic states, become unfavorable once the exciton is in the acceptor phase. In terms of the rate model, this corresponds to an increase of the degeneracy ratio $g_{\mathrm{CT}}/g_{\mathrm{LE}}$ by a factor of \num{2}. Note that this effective loss of excitonic state does not change $g_{\mathrm{CT}}$, because CT states are assumed to only exist near donor--acceptor interfaces and all interfaces are still accessible from the acceptor phase. By rearranging Eq.\,\ref{IQErateSolution}, having substituted Eq.\,\ref{probratio} to eliminate $k_{\mathrm{21}}$, we get
\begin{equation}
	\Delta E_{\mathrm{LE-CT}} = - k_{\mathrm{B}}T \left(\ln \left(\frac{g_{\mathrm{CT}}}{g_{\mathrm{LE}}}\right) + f(k,\,\mathrm{IQE})\right),
	\label{inverseEQ}
\end{equation}
where $f(k,\,\mathrm{IQE})$ is some function of all rate constants of the model as well as the IQE. With Eq.\,\ref{inverseEQ} it is obvious that doubling $g_{\mathrm{CT}}/g_{\mathrm{LE}}$ leads to a shift of the IQE curve of $-\ln(2)k_{\mathrm{B}}T\approx \SI{-17.5}{\milli\eV}$ at $T=\SI{293}{\kelvin}$.

\subsection{Gaussian energetic disorder}

\begin{figure*}
\includegraphics[width=1\textwidth]{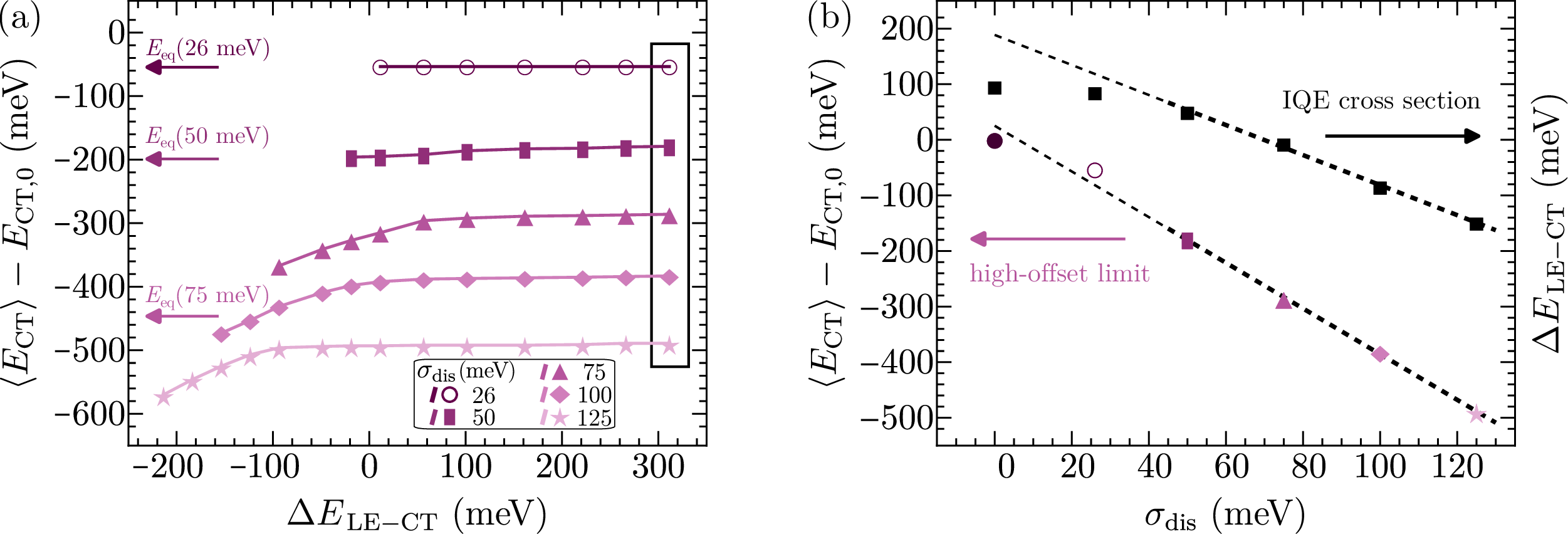}
\caption{(a) kMC data for the average energy $\langle E_{\mathrm{CT}} \rangle$ of the occupied density of states (DOS) of the steady state charge-transfer (CT) population relative to the center of the original DOS $E_{\mathrm{CT,0}}$ as a function of the exciton--CT offset \mbox{$\Delta E_{\mathrm{LE-CT}}=\Delta E_{\mathrm{LE}}-E_{\mathrm{CT,0}}$}. Since the energy distributions in the underlying data are not perfectly symmetrical in all cases, we used the respective peak positions to approximate the averages. All data plotted originates from the same simulation runs that produced the data in Fig.\,\figref{disresults}. The equilibrium energies $E_{\mathrm{eq}}$ (see Eq.\,\ref{equilibrium}) for three disorder strengths are indicated by arrows on the left. All solid lines are merely guides to the eye. (b) The values of the high-offset limit from (a), marked by a black rectangle, and the cross section data from Fig.\,\figref[(a)]{disresults} are plotted as a function of the standard deviation $\sigma_{\mathrm{dis}}$ of the Gaussian used for modeling energetic disorder. Data points corresponding to high $\sigma_{\mathrm{dis}}$ are fitted with a linear function (thick-dashed lines) in each case. The thin-dashed lines are just an extension of the fit serving as a visual guide. As expected from Eq.\,\ref{thermalization}, a linear dependence emerges due to hopping relaxation in a Gaussian DOS. The energetic relaxation of the CT states creates an effective energetic offset, which in turn yields a significant driving force for charge separation.}
\label{fig:disLinFit}
\end{figure*}

The same logic holds for kMC data including Gaussian disorder. For this reason, we limit the following discussion of disorder effects to scenario 1, i.\,e., the LUMO--HOMO shift of the acceptor. As discussed before, this yields a constant exciton energy \mbox{$E_{\mathrm{LE}}=E_{\mathrm{LE,don}}=E_{\mathrm{LE,acc}}$}. Since the CT energy in a disordered landscape is technically neither a constant in space nor in time anymore, we define $E_{\mathrm{CT,0}}$ to be the average energy of the density of states (DOS), i.\,e., the peak of the Gaussian  \footnote{Every charge transfer (CT) state is consisting of an electron and a hole on different sites. If the HOMO and LUMO energies of the sites include a random Gaussian shift with standard deviation $\sigma_{\mathrm{dis}}$, the density of states for CT states will be a Gaussian with standard deviation $\sqrt{2}\,\sigma_{\mathrm{dis}}$.}. Accordingly, the exciton--CT offset is now defined as $\Delta E_{\mathrm{LE-CT}}=E_{\mathrm{LE}}-E_{\mathrm{CT,0}}$. When talking about the average energy of the \emph{occupied} DOS at time $t$, we denote it as $\langle E_{\mathrm{CT}}(t) \rangle$. $\langle E_{\mathrm{CT}} \rangle$ is the steady-state limit of $\langle E_{\mathrm{CT}}(t) \rangle$. Fig.\,\figref[(a)]{disresults} shows the effect increasing the energetic disorder $\sigma_{\mathrm{dis}}$ has on the IQE curves. For the same values of $\sigma_{\mathrm{dis}}$, Fig.\,\figref[(b)]{disresults} plots the relative steady-state populations of both LE and CT states. Going forward, we split the discussion of the effects of the Gaussian disorder into two categories. One is the classic high-offset limit. The other consists of low-offsets relevant for state-of-the-art systems.

The trends observed for high offsets coincide nicely with the literature, e.\,g., the studies of Wojcik et al. \cite{wojcikElucidatingRoleDisorder2022}. 
Firstly, increasing the energetic disorder has a detrimental effect on the IQE at $\Delta E_{\mathrm{LE-CT}}\gg 0$ in systems with an efficient separation to begin with. Excitons are rapidly dissociated into CT states. Charge back-transfer is not probable in this regime, so CT states have to either recombine to the ground state or separate. If charge separation in a homogeneous energetic landscape, i.\,e., without disorder is efficient, the unfavorable high and low energy sites for electrons and holes, respectively, that are introduced by the disorder will hinder separation. The average charge mobility is declining and CT recombination increases. Additionally, the electron/hole of the CT pair will jump locally, preferably to neighboring sites of lower/higher energy. This thermalization in the disordered DOS further increases the trapping \cite{melianasPhotogeneratedChargeTransport2019}. In the present simulation, this leads to a loss of over \SI{25}{\percent} efficiency of charge separation and an approximately equal increase of CT states present in the steady-state at $\sigma_{\mathrm{dis}}\approx3\,k_{\mathrm{B}}T$. Eventually though, high values of $\sigma_{\mathrm{dis}}$ will cause the high-offset IQE to increase again \cite{albrechtYieldGeminatePair1995}. In our case, IQE increases by about \SI{5}{\percent} when going from $\sigma_{\mathrm{dis}}=\SI{75}{\milli\eV}$ to \SI{125}{\milli\eV}. Interestingly enough, this opposite trend can be explained by the same two phenomena. A disordered energy landscape will occasionally provide a more suitable pathway for separation than the average might allow. Furthermore, if the CT charges already managed to separate by a few nanometers, the energetic relaxation into the tail of the DOS effectively prevents them from approaching each other again and delaying recombination. In other words, the energetic loss of thermalization counteracts the gain necessary to leave the Coulomb well \cite{albrechtYieldGeminatePair1995,melianasPhotogeneratedChargeTransport2019}. In order to illustrate the CT thermalization, we plotted $\langle E_{\mathrm{CT}} \rangle$ relative to the center of the DOS in Fig.\,\figref[(a)]{disLinFit}. The high-offset values marked by the black rectangle are displayed as a function of $\sigma_{\mathrm{dis}}$ in Fig.\,\figref[(b)]{disLinFit}. Analytically, the time-dependence of the average energy of the occupied DOS during thermalization in a Gaussian DOS is approximately proportional to $-\sigma_{\mathrm{dis}}$ \cite{basslerChargeTransportDisordered1993,riesMonteCarlostudyRelaxation1988}:
\begin{equation}
    \langle E_{\mathrm{CT}}(t) \rangle - E_{\mathrm{CT,0}} \propto -\sigma_{\mathrm{dis}}\, \ln(t/t_0).
    \label{thermalization}
\end{equation}
$t_{0}$ is a constant dwell time depending on $\nu_{0,\mathrm{e/h}}$, $\alpha$ and the lattice constant. This drop in energy will stop either once the equilibrium energy $E_{\mathrm{eq}}(\sigma_{\mathrm{dis}})$ is reached \footnotemark[\value{footnote}],
\begin{equation}
    \langle E_{\mathrm{CT}} \rangle = E_{\mathrm{eq}}(\sigma_{\mathrm{dis}}) = E_{\mathrm{CT,0}}-2\,(\sigma^2_{\mathrm{dis}}/k_{\mathrm{B}}T),
    \label{equilibrium}
\end{equation}
or if the charges recombine. $E_{\mathrm{eq}}(\sigma_{\mathrm{dis}})$ is marked in Fig.\,\figref[(a)]{disLinFit} for the first three values of $\sigma_{\mathrm{dis}}$. We see that for the parameters in our simulation, thermal equilibrium of the steady-state CT population is achieved for $\sigma_{\mathrm{dis}}<2\, k_{\mathrm{B}}T$. Since the average CT lifetime is limited and approximately independent of $\sigma_{\mathrm{dis}}$, the average energies $\langle E_{\mathrm{CT}} \rangle$ for $\sigma_{\mathrm{dis}}\geq 2\, k_{\mathrm{B}}T$ should scale linearly with $\sigma_{\mathrm{dis}}$ according to Eq.\,\ref{thermalization}. This is confirmed by the high-offset data in Fig.\,\figref{disLinFit}.

For small $\Delta E_{\mathrm{LE-CT}}$, the data in Fig.\,\figref{disresults} reveals a substantial gain of efficiency with increasing $\sigma_{\mathrm{dis}}$. Depending on $\Delta E_{\mathrm{LE-CT}}$, the magnitude of this boost can even exceed the loss of IQE at high offsets. Although not an exact match, the IQE curves in Fig.\,\figref[(a)]{disresults} reveal a crossover point to the beneficial regime at $\Delta E_{\mathrm{LE-CT}}\approx \SI{130}{\milli\eV}$. As disorder increases, the original sigmoidal shape gets stretched along the $\Delta E_{\mathrm{LE-CT}}$ axis, resulting in the high-offset IQE plateau reaching lower offsets. The cross section of the IQE data along the dashed line is plotted in Fig.\,\figref[(b)]{disLinFit} as a function of $\sigma_{\mathrm{dis}}$. As before, a linear dependence for high disorders is evident and connects well with the underlying mechanism of relaxation in the Gaussian DOS. Repopulation and subsequent recombination of excitons are the main source of IQE loss at low offsets. This is directly visualized by the steeply increasing exciton populations in Fig.\,\figref[(b)]{disresults}. As $\Delta E_{\mathrm{LE-CT}}$ gets smaller, the energetic barrier for a charge back transfer decreases. Because the average lifetime of excitons is much smaller than the one of CT states, this repopulation yields increased recombination losses. The thermalization of charge carriers counteracts this trend by shifting $\langle E_{\mathrm{CT}} \rangle$ well below $E_{\mathrm{CT,0}}$. The resulting \emph{effective} offset $E_{\mathrm{LE}}-\langle E_{\mathrm{CT}} \rangle$ can be interpreted as a substantial driving force, just as in the high-offset limit discussed before.

Interestingly, the thermalization data in Fig.\,\figref[(a)]{disresults} seems to imply an additional energetic relaxation of the CT population starting at decreasingly smaller offsets with rising $\sigma_{\mathrm{dis}}$. We interpret this as a statistical effect rather than a real shift due to carriers reaching lower energies. Firstly, while the CT DOS is sampled evenly by dissociating excitons at high offsets, the closer $E_{\mathrm{CT,0}}$ and $E_{\mathrm{LE}}$ get, the lower the average of initially occupied CT states becomes. Furthermore, as exciton repopulation becomes feasible at low offsets, CT states with energies close to $E_{\mathrm{LE}}$ are more likely to recombine as excitons, which also shifts the average CT decay energy downward.

Finally, if $E_{\mathrm{LE}}\ll E_{\mathrm{CT}}$, then $\Delta E_{\mathrm{LE-CT}}\ll 0$ and exciton dissociation into CT states at interface sites will become a strictly thermally-activated process. The majority of generated excitons will decay without separating at least once. Nonetheless, interface sites with CT energies in the low-energy tail of the DOS will allow separation to be more likely to occur. In consequence, increasing $\sigma_{\mathrm{dis}}$ will increase IQE. Because the number of interface sites an exciton can reach during its lifetime is limited and the density of tail states sinks rapidly, IQE will approach \SI{0}{\percent} for all levels of disorder. In this regime, since charge separation is limited by the low number of energetically favorable CT states, exciton dissociation becomes a diffusion-limited process.

\section{Conclusion}

Low-offset material blends are an essential part in state-of-art organic solar cell (OSC) research. They allow for higher open-circuit voltages, but challenge the traditional paradigm of high driving-force charge separation. Although first-order rate equations are able to capture the essence of low-offset equilibria between excitons and charge-transfer states, they are not able to model complex effects such as morphology and disorder. Kinetic Monte-Carlo (kMC) simulation allows multi-scale modeling of OSCs at the cost of high computation time. In this work, we connected the predictions of the steady-state rate model to kMC data, both validating the employed kMC implementation and showing the limits of the analytical model.

When ignoring disorder effects, rate model and kMC produce nearly identical effects. Remarkably, the differences between the two scenarios of accomplishing the exciton--CT offset in the kMC data can be explained entirely by the ratio of degeneracies of exciton and CT states in the rate model. The corresponding offset-shift observed in the data when changing the scenario allowed us to limit the disorder study to one of the scenarios. Including Gaussian disorder in the kMC revealed detrimental as well as beneficial effects depending on the energetic offset between exciton and CT state. For high offsets, where exciton dissociation and CT separation are dominant, disorder will mostly decrease separation efficiency. At large disorders, a slight upward trend in efficiency appears, which is consistent with the literature on this matter. Medium and low offsets exhibit an increase of efficiency with rising energetic disorder. This boost can be significant, even exceeding the IQE loss observed at high offsets. As a consequence, substantial charge separation yield can be observed at lower and lower offsets with increasing disorder. In this regime, charge carrier thermalization in the Gaussian density of states creates an effective energetic offset, which is much larger than the one without disorder. This results in a net driving force for charge separation that increases with energetic disorder.

From a practical point of view the present investigation showcases that the range of applicable low-offset material combinations for high-efficiency solar cells can be extended to small and even negative offsets by controlling the energetic disorder in the system. Finally, the possibilities for future kMC studies of low-offset systems are evident. We plan to expand the current simulation setup to include mirror charge effects regarding the contacts in order to gain deeper understanding of the intricacies of low-offset organic photovoltaics.

\bibliography{references.bib}

\end{document}